# LV-CARE-Diff: A Conditional Anatomy-Aware Diffusion Model for Left Ventricular Shape Reconstruction and Function Quantification from Ultra-Sparse Cine Slices

Xinwang Li[1#], Yu Lian[1#], Bowei Liu[2], Yifei Jiang[1], Jingjing Xiao[3], Haiyan Ding[2], Xiangchuang Kong[4,5,6], Rui Guo[1*]

[1]School of Medical Science and Engineering, Beijing Institute of Technology, Beijing, China

[2]Center for Biomedical Imaging Research, School of Biomedical Engineering, Tsinghua University, Beijing, China

[3]Bio-Med Informatics Research Centre and Clinical Research Centre, Xinqiao Hospital, Army Medical University, Chongqing, China

[4]Department of Radiology, Union Hospital, Tongji Medical College, Huazhong University of Science and Technology, Wuhan, China

[5]Hubei Provincial Clinical Research Center for Precision Radiology & Interventional Medicine, Wuhan, China

[6]Hubei Province Key Laboratory of Molecular Imaging, Wuhan, China

---

[#]**Xinwang Li** and **Yu Lian** contributed equally to this study and share co-first authorship.

**Address for correspondence**:
**Rui Guo**
E-mail: ruiguo.ok@163.com
School of Medical Science and Engineering, Beijing Institute of Technology, Beijing, China

**Abstract(<250 words)**

Left ventricular functional quantification is an essential examination and is routinely performed using cardiovascular magnetic resonance (CMR) cine imaging. However, conventional CMR cine protocols require the acquisition of multiple short-axis (SAX) slices to cover the entire left ventricle (LV) along with two long-axis (LAX) slices, which is time-consuming and places a considerable burden on patients who are unable to sustain repeated breath-holds, limiting its suitability for large-scale early screening. In this study, a Conditional Anatomy-Aware Diffusion Model (LV-CARE-Diff) was developed using a coarse-to-fine strategy to reconstruct the complete LV shape from ultra-sparse cine slices, namely three short-axis and two long-axis slices, with the aim of accelerating CMR cine examination. LV-CARE-Diff employs a 3DUNet to generate a coarse initial shape, which is subsequently refined through a residual diffusion model. A condition-guided input incorporating imaging plane orientation and positional metadata was constructed to enable spatial awareness, and a multi-objective training strategy jointly supervising shape, function, and anatomy was incorporated to guide high-fidelity reconstruction. LV-CARE-Diff was compared against a standalone 3DUNet, a standalone diffusion model, and a 3D UNet with diffusion-based refinement. Testing results indicated that complete LV shape could be robustly reconstructed by all deep learning models, with the highest reconstruction performance achieved by the proposed LV-CARE-Diff. Deep learning models reconstructing LV shape from sparse cine slices preserved 96% of functional quantification accuracy while reducing imaging time by 73%. The LV-CARE-Diff framework established in this study enables ultra-sparse cine acquisition to shorten CMR examination duration without sacrificing quantitative functional accuracy.

**Highlights (3 bullet points)**

- A Conditional Anatomy-Aware Diffusion Model was developed to reconstruct the complete left ventricular shape from three short-axis and two long-axis cine slices, providing a deep learning solution for accelerating cardiovascular magnetic resonance cine examination.
- A coarse-to-fine reconstruction strategy was adopted, in which a 3D convolutional neural network generates an initial coarse left ventricular shape that is subsequently refined by a residual diffusion model to yield the final reconstruction.
- A condition-guided spatial input encoding the observed slices was combined with a multi-objective training strategy jointly supervising shape, function, and anatomy to drive high-fidelity left ventricular shape reconstruction.

# INTRODUCTION

Heart disease remains the leading cause of death worldwide. In the United States, approximately 6.7 million adults over the age of 20 are living with heart failure (HF), and prevalence is projected to rise to 8.5 million by 2030. The lifetime risk of developing HF has reached 24%, meaning nearly one in four individuals will be affected over their lifetime (1). In China, the prevalence stands at approximately 1,000 per 100,000 individuals (2). Cardiovascular magnetic resonance (CMR) cine imaging is the gold-standard technique for assessing wall motion abnormalities, cardiac morphology, and ventricular function, playing a pivotal role in the diagnosis, risk stratification, and management of patients with reduced ejection fraction (3,4). **Compared with other imaging modalities, CMR cine offers superior reliability for detecting early subclinical dysfunction, particularly in high-risk populations such as patients with diabetes, hypertension, or a history of chemotherapy, and is therefore well suited to long-term longitudinal monitoring (3,5,6).**

In clinical practice, two-dimensional (2D) retrospective electrocardiogram (ECG)-gated segmented cine is the most widely used technique and is performed in virtually every CMR examination (7,8). To achieve adequate spatial and temporal resolution, data for a single 2D slice are acquired over multiple cardiac cycles during a breath-hold. For complete left ventricular (LV) coverage, a dense stack of 10-15 short-axis (SAX) slices together with two long-axis (LAX) views is required, resulting in 8-10 minutes of acquisition time within an approximately 40-minute routine examination (9,10). This substantial scan time reduces overall efficiency, increases cost and patient burden, and consequently impedes large-scale early screening. Although the breath-hold duration per slice is generally tolerable for most patients, repeated breath-holding can be demanding for elderly individuals or those with respiratory comorbidities (11). **There is therefore a clear clinical need for methods that can shorten cine acquisition and, by extension, the entire routine CMR examination, in order to reduce patient burden (12) and to enable earlier screening (4).**

On the imaging acceleration front, single-beat cine has been explored and validated alongside advances in MRI reconstruction techniques (13-15). In contrast to segmented cine, single-beat cine captures all cardiac phases within a single heartbeat, enabling complete ventricular coverage within a single breath-hold or even under free-breathing conditions. However, the aggressive undersampling required inevitably compromises spatial and temporal resolution relative to conventional multi-beat acquisitions. Residual aliasing artifacts and blurring can impair the delineation of fine myocardial structures and the accurate assessment of LV function (13-15). **Consequently, broad clinical adoption of single-beat cine still requires advances in reconstruction algorithms to close the performance gap with conventional segmented cine.**

An alternative strategy is to reduce the number of acquired slices, thereby preserving both spatial and temporal resolution at the cost of reduced through-plane coverage. Previous studies have demonstrated that LV mass and LV ejection fraction can be estimated from cine of a single long-axis (LAX) slice (16-18). However, despite showing good accuracy, the adopted models remain geometrically simple, and full reconstruction of LV shape cannot be achieved. To date, this approach has not been widely adopted in clinical practice. Kuoy et al. and Nguyen et al. demonstrated the feasibility of using three short-axis (SAX) slices, primarily motivated by alleviating the segmentation burden. Although satisfactory agreement was reported between reduced-slice (3 slices) and full-slice (~10 slices) acquisitions in LV functional quantification, both studies relied on conventional algorithms provided by commercial software, without investigating dedicated reconstruction or interpolation algorithms or leveraging the additional spatial context afforded by two-chamber and four-chamber LAX views (19,20). In our previous work, a 3D convolutional neural network was employed to reconstruct LV geometry from sparse short-axis slices conditioned on two-chamber (2ch) and four-chamber (4ch) LAX slices (21), providing an initial demonstration of the feasibility of reconstructing LV shape and function from sparse cine views. However, the model adopted in that work was limited in representational capacity. Meanwhile, various learning-based methods have been proposed to reconstruct LV shape accurately from cine short-axis and long-axis slices (22-34). Among these, Muffoletto et al. proposed a neural implicit representation approach to reconstruct LV shape from as few as half the standard short-axis slices (22), while the remaining studies largely focused on full-stack inputs and did not attempt to reconstruct LV shape from ultra-sparse cine acquisitions. Although these methods enable accurate, automatic, end-to-end reconstruction of the 3D left ventricular shape from cine images, they still rely on a complete slice stack as input, and several approaches require large training datasets (22,28,29). **Taken together, these gaps motivate further investigation into advanced deep learning reconstruction techniques, systematic assessment of agreement with conventional dense-slice acquisitions, and evaluation of clinical feasibility in patient cohorts.**

We hypothesize that conventional slice-dense cardiac cine acquisition can be substantially accelerated through ultra-sparse slice sampling, and that a deep learning-based reconstruction framework can accurately recover left ventricular volumetric and functional indices from such sparse inputs, ultimately achieving clinically meaningful reductions in scan time without compromising diagnostic quality. The main contributions of this study are:

1. A Conditional Anatomy-Aware Diffusion Model (LV-CARE-Diff) was developed to reconstruct the complete left ventricular shape from as few as three ultra-sparse SAX and two LAX cine slices, establishing a deep learning-based framework for accelerated cardiac magnetic resonance cine examination.

2. A coarse-to-fine reconstruction strategy was explored for LV shape reconstruction from ultra-sparse cine slices, in which a 3D convolutional neural network first generated an initial coarse left ventricular shape, subsequently refined by a residual diffusion model to yield the final anatomically accurate reconstruction.

3. A condition-guided input encoding imaging plane orientation and positional metadata was constructed to enable spatial awareness, complemented by a multi-objective training strategy jointly supervising shape fidelity, functional consistency, and anatomical plausibility to guide high-fidelity left ventricular shape prediction.

4. The robustness of LV-CARE-Diff was systematically evaluated with respect to several confounding factors, including cardiac phase, number and location of input SAX slices, and the absence of LAX slice inputs.

## Methods

### *LV-CARE-Diff Design*

**Figure 1** illustrates the overall framework of this study. The objective is to recover the complete LV shape from an ultra-sparse input consisting of 3 SAX slices and 2 LAX slices. This reconstruction problem can be formulated as: $Y=f(X)+n$, where $Y=\{Y_{Endo}, Y_{Epi}\}$ denotes the ground-truth LV shapes comprising the endocardial and epicardial surfaces, $X$ represents the sparse input slices, and $n$ denotes noise. Here, $Y_{Endo}$, $Y_{Epi}$, $n \in R^{W\times H\times D}$, and W×H×D denotes the spatial dimensions of the output volume.

When dense SAX slices spanning from apex to base are available alongside LAX slices, $f(\cdot)$ can be realised by conventional approaches such as shape-based interpolation, statistical shape models, or atlas-based methods (35,36). However, given only 3 ultra-sparse cine SAX and two LAX slices, this reconstruction problem becomes severely ill-posed, and the complete LV shape can no longer be reliably recovered by such means. To address this, we proposed a deep learning approach, LV-CARE-Diff, to approximate $f(\cdot)$, with the objective of reconstructing the complete LV shape while preserving accurate functional quantification.

As shown in **Figure 1A**, LV-CARE-Diff comprises two subnetworks, LVRecNet and DiffResNet, following a coarse-to-fine strategy. A six-channel spatial condition tensor $X^{'} \in R^{W \times H \times D \times 6}$ is constructed from the sparse observations $X$ to serve as condition-guided input for both networks. LVRecNet takes $X^{'}$ as input and recovers a coarse LV shape $\hat{Y}^{C}$, which is subsequently refined iteratively by DiffResNet. This process can be expressed as $\hat{Y}^{F} = \hat{Y}^{C} + \eta \tanh(\hat{R})$, where $\hat{Y}^{F}$ represents the final output, $\hat{R}$ is the residual for refining $\hat{Y}^{C}$, $\eta$ is a hyperparameter balancing the contribution of the predicted residual relative to the coarse estimate, and the *tanh* function constrains the magnitude of the predicted residual.

*Condition-guided Inputs*

As shown in **Figure 1B**, the six channels of X' are designed as follows: one epicardial mask ($\mathrm{Mask}_{\mathrm{Epi}}$), one endocardial mask ($\mathrm{Mask}_{\mathrm{Endo}}$), one observed mask corresponding to the SAX stack ($\mathrm{Mask}_{\mathrm{observed,\ SAX}}$), two observed masks corresponding to the two LAX slices ($\mathrm{Mask}_{\mathrm{observed,2ch}}$ and $\mathrm{Mask}_{\mathrm{observed,4ch}}$), and one z-map. $\mathrm{Mask}_{\mathrm{Epi}}$ and $\mathrm{Mask}_{\mathrm{Endo}}$ are constructed by mapping the segmented epicardial and endocardial contours from both SAX and LAX cine images into the 3D volume, using the image-position-patient, image-orientation-patient, slice location, and pixel spacing parameters extracted from the DICOM. Each $\mathrm{Mask}_{\mathrm{observed}}$ encodes the acquisition status of the corresponding plane: a value of one denotes a voxel location belonging to an observed-plane slice, while zero denotes a location outside that plane. These observed-plane prompts provide spatial acquisition cues to guide the network and do not constrain the output at the observed locations. The z-map is a depth-coordinate map that varies linearly from -1 to 1 along the apex-to-base axis, encoding the relative position of each slice within the LV volume.

*Coarse LV Shape Reconstruction*

LVRecNet is designed as a 3DUNet comprising two downsampling blocks, two upsampling blocks, and a bottleneck block, with skip connections concatenating encoder and decoder feature maps at each resolution level. Each block consists of two 3D convolutional layers (Conv3D), each followed by a ReLU activation function. Max-pooling (MaxPool3D) and transposed convolution (ConvTranspose3D) operations are adopted for downsampling and upsampling, respectively. At the end, a 1×1×1 convolutional layer followed by a sigmoid activation function is applied to produce the coarse LV shape prediction $\widehat{Y}^{C}$.

*Residual Diffusion Refinement*

As shown in **Figure 1C**, DiffResNet is a time-conditioned 3D residual U-Net that shares the encoder-decoder topology of LVRecNet but replaces the standard convolutional blocks with residual blocks, and adopts learned strided convolutions in place of max-pooling and transposed convolutions for downsampling and upsampling, respectively.

The diffusion timestep $t$ is first encoded via a 128-dimensional sinusoidal positional embedding, then processed through a multilayer perceptron to produce a time embedding $E_t$, which is injected into each residual block to enable the network to modulate the network's denoising behavior according to the noise level at each timestep.

For a given input feature tensor $F$ and time embedding $E_t$, each residual block (ResBlock) is computed as: $ResBlock(F, E_t) = Conv3D\left(SiLU\left(GN\left(Conv3D\left(SiLU\left(GN(F)\right)\right) + W_t E_t\right)\right)\right) + W_s F$, where SiLU denotes the Sigmoid Linear Unit activation, GN denotes group normalization, $W_t$ is a learned linear projection of $E_t$ onto the feature space, and $W_s$ is either an identity mapping or a 1×1×1 convolution used to match the channel dimensionality of the residual branch.

The encoder of DiffResNet progressively increases the channel dimension from 32 to 64 across two downsampling stages, each comprising a ResBlock followed by a stride-2 Conv3D for spatial downsampling. A bottleneck ResBlock then operates at the lowest resolution, producing feature maps with 128 and 64 output channels sequentially. The decoder mirrors the encoder using ConvTranspose3D for upsampling and concatenating the corresponding encoder feature maps at each resolution level. Two independent 1×1×1 convolutional output heads produce the predicted noise $\hat{\varepsilon} \in R^{W \times H \times D \times 2}$ for the endocardial and epicardial shapes separately.

At a given diffusion timestep $t$, the residual $R^t \in R^{W \times H \times D \times 2}$, the condition tensor X', and $\hat{Y}^C$ are concatenated along the channel dimension to form a combined input tensor $F \in ^{W \times H \times D \times 10}$, which is fed into DiffResNet. Here, $\hat{Y}^C$ is kept differentiable throughout the conditioning path. Given the noise $\hat{\varepsilon}$ predicted by DiffResNet, the clean residual is recovered as: $\hat{R}_0^{(t)} = \frac{\hat{R}_t - \sqrt{1-\bar{\alpha}_t}\hat{\varepsilon}}{\sqrt{\bar{\alpha}_t}}$, where $\bar{\alpha}_t$ is the cumulative product coefficient of the diffusion schedule. During inference, the denoising process is executed using a 50-step denoising diffusion implicit model (DDIM) sampling scheme (37).

*Structure, Anatomy, Function, and Diffusion Losses*

Multiple loss functions were employed to constrain the model output, collectively supervising LV structure, anatomy, and volume. We first define a joint Binary Cross-Entropy (BCE) and Dice loss with equal weighting for a paired prediction $\widehat{P}$ and its reference $P^{\mathrm{Ref}}$: $\mathcal{L}_{\mathrm{BD}}(\widehat{P}, P^{Ref}) = 0.5\mathcal{L}_{\mathrm{BCE}}(\widehat{P}, P^{Ref}) + 0.5\mathcal{L}_{\mathrm{Dice}}(\widehat{P}, P^{Ref})$. Here, BCE provides voxel-wise supervision, whereas Dice measures overlap and is less sensitive to class imbalance caused by background dominance.

For each category $q \in \{Endo, Epi\}$, the whole-LV shape loss at diffusion timestep $t$ is defined as $\mathcal{L}_{\mathrm{LV,q}}^{(t)} = \mathcal{L}_{\mathrm{BD}}\left(\widehat{Y}_q^{(t)}, Y_q^{Ref}\right)$. For the acquired SAX slices, the corresponding slices are resampled from $\widehat{Y}_q^{(t)}$ at the locations of the acquired planes, and the joint BCE and Dice loss is calculated and denoted $\mathcal{L}_{\mathrm{SAX,\,q}}^{(t)}$. For the LAX slices, a differentiable signed distance field (SDF) proxy is first computed for the predicted LV shape, denoted *sdf*(·). The image plane of each LAX view $v$ is then resampled from this proxy volume via a projection operator *proj*(·,$v$), yielding a 2D SDF slice for that view. The SmoothL1 loss is then computed between the projected signed-field slices and the corresponding reference SDF slice $\mathrm{SDF}_v^{Ref}$ for LAX consistency. The whole-volume and SAX terms supervise global and observed-plane occupancy, whereas the projected LAX term checks boundary position and signed inside/outside geometry on the acquired long-axis views. Let $\mathcal{V}_i = \{1, 2\}$ denote the set of valid LAX slices, where 1 and 2 correspond to the 2ch and 4ch acquisitions, respectively. The LAX consistency loss is then defined as:

$$\mathcal{L}_{\mathrm{LAX},q}^{(t)} = \frac{1}{|V_i|}\sum_{V_i} \mathrm{SmoothL1}\left(\mathrm{proj}\left(sdf\left(\hat{Y}_q^{(t)}\right), V_i\right), \mathrm{SDF}_{V_i}^{Ref}\right)$$

If no LAX slices are available, $\mathcal{L}_{\mathrm{LAX},q}^{(t)}$ is set to 0.

The combined shape loss for each category $q$ is defined as:

$$\mathcal{L}_{Shape,\ q}^{(t)} = 1.0\mathcal{L}_{LV,q}^{(t)} + 0.5\mathcal{L}_{\mathrm{SAX,q}}^{(t)} + 0.3\mathcal{L}_{\mathrm{LAX,q}}^{(t)}$$

Because endocardial errors directly affect cavity volume estimates, the full left ventricular shape loss assigns a higher weight to the endocardial region and is defined as:

$$\mathcal{L}_{\mathrm{LVShape}} = 1.5\mathcal{L}_{Shape,\mathrm{Endo}}^{(\mathrm{t})} + \mathcal{L}_{Shape,\mathrm{Epi}}^{(\mathrm{t})}.$$

The LV myocardium is additionally constrained using the joint BCE and Dice loss, denoted $\mathcal{L}_{\mathrm{Myo}}^{(t)}$. The overall structure loss, combining the refined prediction $\hat{Y}^{(t)}$, the myocardium term, and the coarse prediction $\hat{Y}^{C}$ from LVRecNet, is defined as

$$\mathcal{L}_{\mathrm{Structure}} = \mathcal{L}_{LVShape,\hat{Y}^{(t)}}^{(t)} + 0.2\,\mathcal{L}_{\mathrm{Myo}}^{(t)} + 0.8\mathcal{L}_{LVShape,\hat{Y}^{C}}^{(t)}$$

*Anatomy loss:* to enforce the geometric constraint that the endocardial shape lies entirely within the epicardial shape, the anatomy loss is defined as

$$\mathcal{L}_{\mathrm{Anatomy}} = \mathrm{mean}\left\{\mathrm{ReLU}\left[\hat{Y}_{Endo}^{(t)} - \hat{Y}_{Epi}^{(t)}\right]\right\}.$$

*Function loss:* to supervise LV function, a volume loss is computed between the predicted volume and reference volume:

$$\mathcal{L}_{\mathrm{Vol}} = \mathrm{SmoothL1}\left(\hat{V}, V^{Ref}\right).$$

*Diffusion loss*: *the diffusion noise loss was defined as:*

$$\mathcal{L}_{\mathrm{Diff}} = \frac{1}{2}\left(\mathrm{MSE}(\hat{\varepsilon}_{Endo}, \varepsilon_{\mathrm{Endo}}) + \mathrm{MSE}\left(\hat{\varepsilon}_{Epi}, \varepsilon_{\mathrm{Epi}}\right)\right).$$

*Anchor loss:* To prevent the coarse LV prediction from being unexpectedly deformed by the diffusion residual refinement, an anchor loss penalizes large deviations between the refined output and the coarse prediction:

$$\mathcal{L}_{\mathrm{Anchor}} = \mathrm{mean}\left|\hat{Y}_{LV,q}^{(t)}, Y_{LV,q}^{C}\right|$$

Overall, the total loss at diffusion step *t*, used to update LV-CARE-Diff, is a weighted sum of all the above terms:

$$\mathcal{L}_{\mathrm{total}} = \mathcal{L}_{\mathrm{Sturcture}} + 0.05\mathcal{L}_{Anatomy} + 0.10\mathcal{L}_{\mathrm{Vol}} + \mathcal{L}_{Diff} + 0.08\mathcal{L}_{\mathrm{Anchor}}$$

*MR dataset and Reference Construction*

The cine dataset comprised 240 subjects (128 males, 32.9 ± 13.6 years), of whom 153 were healthy volunteers (71 males, 26.1 ± 5.6 years) and 87 were patients (57 males, 44.8 ± 15.3 years). Healthy volunteers were scanned at Tsinghua University, while patient data were acquired at Wuhan Union Hospital. All MRI studies were approved by the local Institutional Review Board, and written informed consent was obtained from all participants prior to scanning. All healthy volunteers were scanned at 3 T, while 45 patients were scanned at 3 T and the remaining 42 at 1.5 T. Each subject was imaged with 10-12 SAX slices covering the left ventricle, with a slice thickness of ~8 mm and a slice gap of 2 mm; the number of slices was determined by individual heart size. In addition, one 4ch and one 2ch views were acquired per subject. The balanced steady-state free precession (bSSFP) cine sequence acquired approximately 15 phase-encoding lines per segment per cardiac phase using an ECG-triggered readout, with two-fold Compressed SENSE acceleration. Other imaging parameters were consistent with those adopted in clinical practice.

Epicardial and endocardial contours were delineated slice by slice for the end-diastolic and end-systolic phases using CVI42 (Circle Cardiovascular Imaging, Calgary, Canada), with a CMR expert manually reviewing and correcting any cases in which automatic segmentation was unsatisfactory. For each subject, the segmented contours from all SAX and LAX slices were projected into a 3D volumetric matrix using the MRI acquisition parameters, namely pixel spacing, slice location, image position, and image orientation, to construct a point-cloud representation of the reference LV shape $Y_q^{Ref}$ at a target resolution of $1\times1\times1$ mm$^3$. The SAX plane was aligned parallel to the *xy*-plane, and the slices were ordered along the z-direction from apex to base. The sparse volumetric representation of $Y_q^{Ref}$ was subsequently reconstructed using a case-specific radial-field and continuous-SDF method, with the annotated SAX masks reinserted at the corresponding z-locations during the intermediate reconstruction. Anisotropic SDF regularization with volume-preserving binarization was then applied, followed by largest-component retention, Endo-inside-Epi correction, and terminal open-base repair, completing the construction of $Y_q^{Ref}$.

The full cohort of 240 subjects with successfully reconstructed reference LV shapes was partitioned into training (106 healthy volunteers and 62 patients), validation (22 healthy volunteers and 14 patients), and testing (25 healthy volunteers and 11 patients) sets, corresponding to 70%, 15%, and 15% of the total, respectively. Of the 11 patients in the testing set, 8 were male, the mean age was 41.5 ± 16.5 years, and 7 were imaged at 1.5 T.

*Implementation and Training*

All models were developed using the PyTorch library (Meta Platforms, Menlo Park, CA, USA) and implemented on a dedicated deep-learning workstation equipped with dual Intel Xeon Gold 6226R CPUs (32 cores each), one NVIDIA RTX 4090 graphics processing unit (24 GB memory), and 512 GB RAM.

The two subnetworks of LV-CARE-Diff were trained jointly using the AdamW optimizer, with momentum coefficients $\beta_1$ =0.9 and $\beta_2$ =0.999, a base learning rate of $2\times10^{-4}$, and a weight decay coefficient of $1\times10^{-4}$. A linear learning rate warmup strategy was applied over the first 5 training epochs, followed by a cosine annealing learning rate schedule. Training was conducted for 100 epochs with a batch size of 1 and gradient clipping was applied with a maximum norm threshold of 0.25. For each training batch, paired end-diastolic and end-systolic phase data were fed into LV-CARE-Diff to update the model parameters. Within each training epoch, the number of input SAX slices was randomly sampled from 3, 4, or 5 slices per volume, and the slice locations were randomly perturbed to adjacent positions to improve model robustness.

For DiffResNet, η is fixed as 0.2. The diffusion process adopts T=1000 steps, with the noise variance $\beta_t$ increasing linearly from $10^{-4}$ to 0.02. The signal retention coefficient is defined as $\alpha_t = 1-\beta_t$, with cumulative product $\bar{\alpha}_t = \prod_{i=1}^{t} \alpha_i$. The forward noising process at an arbitrary timestep $t$, randomly sampled from [1, T], is formulated as: $R_t = \sqrt{\overline{\alpha}_t} R_0 + \sqrt{1-\overline{\alpha}_t} \varepsilon, \varepsilon \sim \mathcal{N}(0, I)$.

*Testing studies*

As no established method exists for reconstructing LV shape from ultra-sparse slice acquisitions, LV-CARE-Diff was compared against three baseline models: a standalone 3DUNet, a standalone diffusion model, and a two-stage model in which the 3DUNet was first trained and frozen, followed by training a residual diffusion model on its output (3DUnet/Diffusion). These baselines represent direct regression, direct diffusion generation, and sequential coarse-to-fine refinement. Respectively, to demonstrate the effectiveness of the coarse-to-fine end-to-end training strategy adopted in LV-CARE-Diff differs from the third model in jointly optimizing the coarse and diffusion branches. The network architecture, implementation, training parameters, and loss functions were kept identical across all methods, and all were tested using three SAX slices.

For LV-CARE-Diff, we further analyzed the impact of cardiac phase (end-diastolic versus end-systolic), subject cohort (healthy volunteers versus patients), SAX slice number (increasing from 3 to 5 slices), SAX slice location (replacing the pre-calculated targeted slices with adjacent slices), and LAX slice availability (omitting one LAX view) on reconstruction performance.

*Quantitative metrics*

Dice coefficient and 95% Hausdorff distance (HD) were used to evaluate LV shape reconstruction performance, while mean absolute error (MAE) was used to assess LV functional index accuracy. Agreement in LV functional indices, encompassing end-diastolic volume (EDV), end-systolic volume (ESV), stroke volume (SV), and left ventricular ejection fraction (EF), between shapes reconstructed from ultra-sparse slices and those derived from the full SAX stack was examined through linear regression and Bland-Altman analysis. A paired Student's t-test was performed to assess statistically significant differences between each deep-learning model and the reference standard for each functional index, with statistical significance defined as $P < 0.05$. Mean difference, relative error, and t-test results are reported in the Supplementary Material.

## Results

### *1. Feasibility Demonstration*

All deep-learning models were trained successfully and performed LV shape reconstruction for all subjects; no cases were excluded from analysis. Code, data, and results are available on GitHub (https://github.com/CMRatBIT/LV-CARE-Diff). In the testing dataset, the mean number of acquired SAX slices per subject was 11.22 (range: 8-15). By reducing the input to three SAX slices, a mean acquisition time saving of approximately 73% could be achieved.

In **Figure 2**, both the end-diastolic and end-systolic LV shapes reconstructed from three SAX slices by the proposed LV-CARE-Diff closely matched those reconstructed from all SAX slices. LV-CARE-Diff also robustly reconstructed the LV shape from three SAX slices for a patient with mildly abnormal LV morphology. For both the healthy volunteer and the patient, the Dice coefficients of the retrospectively resampled slices further confirmed the close agreement between the LV shapes reconstructed from all SAX slices and three SAX slices. These results demonstrate the feasibility of deep learning–based LV shape reconstruction from substantially reduced SAX acquisitions.

*2. LV Recovery Performance Comparison*

**Figures 3** and **4** visually compare LV shape reconstruction across four deep-learning models for two healthy volunteers and two patients. Shapes reconstructed by 3D UNet exhibited spatial distortion and spurious structure generation, particularly in the patients (**Figure 4**), resulting in lower Dice coefficients and less accurate LV functional indices. The standalone diffusion model tended to over-smooth the reconstructed LV shape at the basal region, as observed in the second healthy volunteer and the end-diastolic shapes of the two patients. In contrast, LV shapes reconstructed by both 3DUnet/Diffusion and LV-CARE-Diff closely matched the reference standard, yielding comparable Dice coefficients and LV functional indices.

As shown in **Figure 5A-C** and **Table 1**, LV-CARE-Diff achieved the highest Dice coefficients across all methods for both epicardial and endocardial contours at the end-diastolic and end-systolic phases. Both 3DUnet and 3DUnet/Diffusion yielded comparable LV shape reconstruction performance, with only minimal differences relative to LV-CARE-Diff. In contrast, the standalone diffusion model produced the lowest Dice coefficients, consistent with the visual results in **Figures 3** and **4**. Correspondingly, as shown in **Figure 5D-E** and **Table 1**, LV-CARE-Diff achieved the smallest HD across all methods, while the standalone diffusion model yielded the largest HD error. Consistent with the Dice coefficient results, 3DUnet and 3DUnet/Diffusion exhibited comparable HD errors.

As shown in **Figure 5G-J**, and **Tables 1** and **S1**, LV-CARE-Diff achieved comparable LV functional index quantification to all other methods across all four indices (EDV: 3.29 ± 5.31 mL, ESV: 2.59 ± 2.85 mL, SV: 2.97 ± 3.35 mL, LVEF: 1.80 ± 1.60%). 3DUnet/Diffusion achieved the lowest MAE for ESV (2.48 ± 2.51 mL) and LVEF (1.65 ± 1.62%), and 3DUnet achieved the lowest MAE for SV (2.76 ± 2.99 mL). As shown in **Figures 6** and **S1**, strong correlations were observed between LV functional indices derived from all SAX slices and those obtained from ultra-sparse 3 SAX slices with deep-learning reconstruction across all methods for LV EDV, ESV, and EF ($r > 0.98$), with the exception of LVEF estimated by 3DUnet and the standalone diffusion model. For SV, all correlation coefficients exceeded 0.94. In the agreement analysis (**Figures 7** and **S1**), 3DUnet/Diffusion achieved minimal bias across all four LV functional indices, while the standalone diffusion model exhibited the largest bias in ESV, SV, and LVEF. The corresponding bias values for LV-CARE-Diff were -2.21 mL for EDV, 0.55 mL for ESV, -2.76 mL for SV, and -1.64% for EF.

*3. Impact of cardiac phase*

In **Tables 2** and **S2**, the difference in reconstruction performance between the end-diastolic and end-systolic cardiac phases was negligible, with no apparent difference except for a slightly higher Dice coefficient in the myocardium at the end-systolic phase.

*4. Impact of Subject cohorts*

**Table 3** and **S3** examine the difference in performance between healthy volunteers and patients. LV-CARE-Diff achieved higher accuracy for both LV shape reconstruction and functional index quantification in healthy volunteers than in patients. Although LV-CARE-Diff exhibited slightly higher prediction error in EDV and ESV for patients, the accuracy in LVEF remained well preserved without significant MAE (1.71 ± 1.34% for healthy volunteers and 2.01 ± 2.13% for patients).

*5. Impact of LAX slice inputs*

As LAX slices provide the main spatial structural information for LV-CARE-Diff, the model achieved the highest overall performance when both 2ch and 4ch slices were provided as input, as shown in **Tables 4** and **S4**, with the exception of EDV and ESV. In **Table 4**, the reduction in LV shape reconstruction performance when only one LAX view was provided was not significant compared to inputting both views. Similarly, the reduction in LV functional index quantification when only one LAX view was provided was not significant compared to inputting both views, as indicated by the EF MAE of 1.99 ± 1.59% with 4ch input only, 1.81 ± 1.65% with 2ch input only, and 1.80 ± 1.60% with both views.

*6. Impact of SAX Slices and Location Shift*

In **Tables 5** and **S5**, increasing the number of input SAX slices from three to five improved both LV shape reconstruction (Hausdorff distance reduced from 2.0 mm to 1.5 mm) and LV functional index quantification by approximately twofold (1.80±1.60% VS 0.86±0.66%). However, the improvement observed when inputting five SAX slices was not significant compared to inputting four SAX slices (0.91±0.64% vs. 0.86±0.66%).

As shown in **Tables 6** and **S6**, although LV shape reconstruction performance was not significantly affected by random slice location shifts, LV functional index quantification degraded by approximately twofold (1.80±1.60% vs. 3.60±4.06%) when slice locations were randomly shifted.

## Discussion

In this study, we sought to recover LV volumetric shape and functional indices from ultra-sparse SAX cine slices through constructing a conditional anatomy-aware diffusion Model, with the dual objective of enabling rapid LV functional examination and rescuing LV function assessment from incomplete or corrupted SAX datasets. Our results demonstrate that three SAX slices combined with two LAX views can achieve 96% accuracy in LV functional estimation in both healthy volunteers and patients, relative to conventional full-stack acquisition, while reducing scan time by approximately 73%.

In this study, LV-CARE-Diff is designed as a two-subnetwork architecture implementing a coarse-to-fine strategy to reconstruct the complete LV shape. LVRecNet is trained to extract coarse shape features from ultra-sparse cine slices, while DiffResNet progressively refines the coarse left ventricular shape through iterative denoising, combining the complementary strengths of convolutional and diffusion-based modelling. The results demonstrated that this coarse-to-fine architecture outperformed both a standalone 3DUNet and a standalone diffusion model.

The gradient-routing strategy in LV-CARE-Diff was designed to separate target stabilization from end-to-end optimization. When constructing the residual target and the coarse reference used by the anchor loss, the coarse prediction was treated with stop-gradient, preventing these supervisory targets from changing through the same branches being optimized. In contrast, the coarse prediction remained differentiable when used as the DiffResNet condition and in the final residual-fusion path, allowing the refined segmentation losses to update both LVRecNet and DiffResNet. This asymmetric design provides stable residual and anchor references while preserving joint optimization of the coarse and refinement networks, thereby limiting excessive residual deformation without decoupling the two subnetworks.

In this study, the training dataset comprised predominantly healthy volunteer data, resulting in an imbalanced distribution that may bias the model toward learning normal LV morphology. Pathological LV structures, however, often exhibit varying degrees of morphological alteration, becoming irregular and thereby increasing the difficulty of accurate LV shape reconstruction from ultra-sparse slices, or even from a full slice stack (22), which accounts for the moderate performance observed in the patient testing. Furthermore, the extent and location of myocardial remodeling vary considerably across patients (e.g., diffuse dilatation, concentric or apical hypertrophy), adding further complexity (38). Therefore, a larger and more diverse patient dataset encompassing a broad range of cardiac pathologies is required to improve the generalizability of the proposed model.

In general clinical practice, the left ventricle is commonly divided into three equal slabs along the long axis, with the center of each slab imaged by quantitative mapping sequences such as cardiac $T_1$ and $T_2$ mapping (39,40). Based on this widely adopted acquisition protocol, slices at or near the center of each slab were selected as model input. To improve robustness to slice location variability, the target slice of each slab was randomly replaced with an adjacent slice during training. However, statistical analysis revealed that slices located at the slab centers were sampled at a higher frequency during training, introducing a positional bias. Consequently, LV-CARE-Diff remains moderately sensitive to slice location, as shown in **Table 6**, with LVEF estimation error increasing from -3.37 ± 4.19% to -4.91 ± 10.97% when slice positions were randomly shifted. Although this increase in relative error is unlikely to significantly impact clinical diagnosis decisions, it nonetheless warrants further investigation into more advanced training strategies to enhance spatial robustness.

Our experiments indicated that increasing the number of input slices yielded no apparent improvement in performance, likely because the LV approximates a truncated ellipsoidal shape whose primary geometry is well captured by slices from representative slabs, while the apical and basal outflow tract regions can be inferred from LAX slices. Although performance showed a moderate decline with only three SAX slices (Table S5), given that each slice requires an individual breath-hold, three SAX slices represent a reasonable balance between scan efficiency and LV functional measurement accuracy. Theoretically, two LAX slices provide critical apical and basal information for recovering LV shape from ultra-sparse SAX inputs. However, experimental results demonstrated that inputting only one LAX slice still yielded high accuracy in LV functional quantification (**Table S4**). This finding further broadens the applicability of the proposed algorithm, extending its utility to scenarios requiring rapid quantification as well as to cases with incomplete cine stack data.

LV-CARE-Diff is not restricted to specific cardiac phases. Testing results indicated no significant difference in reconstruction performance between the end-diastolic and end-systolic phases. Although validation was performed at these two phases only, consistent with standard clinical practice for LV functional index quantification, the proposed LV-CARE-Diff is in principle applicable to any cardiac phase. The resulting phase-wise LV shape estimates could further be leveraged to construct a smooth and temporally continuous deforming LV shape, which may offer additional utility for comprehensive LV shape reconstruction and motion analysis.

The current LV-CARE-Diff framework takes segmented binary contours as input, rendering it inherently insensitive to image contrast (e.g., GRE vs. bSSFP), field strength (e.g., 1.5T vs. 3T), and scanner vendor (e.g., Siemens vs. Philips), which contributes to improved generalizability across acquisition settings (41). In this study, segmentation was performed using CVI42 software, a widely adopted tool in clinical practice. This preprocessing step can equally be performed using alternative semi-automatic or fully automatic segmentation tools, or manual delineation, without significantly impacting model performance. A dedicated segmentation model incorporating automatic cardiac phase identification could be integrated into the pipeline to further improve end-to-end automation; however, this may introduce segmentation-related errors and potentially reduce generalizability. Further investigation and development in this direction are warranted.

With the proposed LV-CARE-Diff framework, a subject-specific adaptive CMR protocol could be achieved in clinical practice. As our results demonstrate that LV-CARE-Diff performs more accurately in cases with regular cardiac morphology, the necessary number and locations of short-axis slices could be prospectively optimized prior to scanning, guided by the referring physician's clinical assessment, patient history, and prior examination records, thereby minimizing the risk of reduced accuracy in functional evaluation with ultra-sparse slice acquisitions. Additionally, the proposed framework would support an inline adaptive scanning strategy, in which the number of short-axis slices is incrementally increased until the estimated functional indices converge to a stable value, and slice locations are optimized according to the long-axis view, yielding an intelligent and efficient clinical examination.

The current version of LV-CARE-Diff focuses exclusively on LV volumetric reconstruction and does not extend to right ventricular volume estimation. A primary technical constraint is that the LAX slices are oriented exclusively across the left ventricle, providing limited structural information of the right ventricle. As LV functional quantification is of greater clinical priority in the majority of cardiac conditions, this scope was considered appropriate for the present study. However, the applicability of the current framework is consequently restricted in patients with predominant right ventricular pathology, such as pulmonary heart disease, where RV functional assessment is clinically essential (42). Future work should extend LV-CARE-Diff to incorporate right ventricular volumetric reconstruction, including the optimization of slice selection strategies and systematic performance evaluation in RV-related conditions.

Although the proposed method demonstrated that LV function can be accurately estimated from a small number of slices (e.g., three SAX slices and one LAX slice), offering a significant reduction in cine acquisition time, this sparse input protocol inherently limits certain clinical applications. Specifically, applications such as whole-heart tissue displacement mapping, assessment of regional wall motion abnormalities, and myocardial radiomics require a complete cine stack and therefore fall outside the scope of the current framework (43,44). Clinicians should be made aware of these limitations prior to adoption of the proposed method.

Moreover, motion artifacts degrade image quality, compromising the reliability of LV functional quantification and necessitating repeat acquisitions (45,46). Although formal validation is pending, preliminary results suggest that LV-CARE-Diff may enable robust LV functional assessment even when a limited number of slices are mildly affected by motion artifact, potentially obviating the need for repeat acquisitions or patient recall.

Several limitations of the present study should be acknowledged. First, as no existing method has been proposed to reconstruct the LV shape from ultra-sparse cine slices, direct comparison with other state-of-the-art methods was not possible. Second, as this work represents a proof-of-concept study, the number of patients included was limited, and findings should be interpreted accordingly. Third, the study was conducted retrospectively, and further prospective validation will be required to confirm the generalizability of the proposed framework. Fourth, the optimal short-axis slice locations for input were not systematically investigated. Finally, other advanced deep learning techniques were not explored, and their potential to further improve reconstruction performance remains to be investigated.

## Conclusion

This study introduced LV-CARE-Diff, a diffusion-based deep learning framework for reconstructing the complete left ventricular shape from ultra-sparse cine acquisitions, offering a practical pathway to accelerated cardiovascular magnetic resonance examination. LV-CARE-Diff adopts a two-stage coarse-to-fine architecture, incorporating multiple well-designed loss functions to jointly supervise shape fidelity, functional consistency, and anatomical plausibility. Testing results demonstrated that LV-CARE-Diff robustly recovered LV shape from three SAX slices, achieving 96% accuracy in LV functional quantification while reducing acquisition time by approximately 70% compared with conventional dense SAX protocols. Although further clinical validation and optimization are warranted, these initial results suggest that LV-CARE-Diff represents a viable solution for streamlined cine workflows and, more broadly, a robust framework for salvaging quantitative LV assessments from incomplete acquisitions.

## Figure and Table Captions

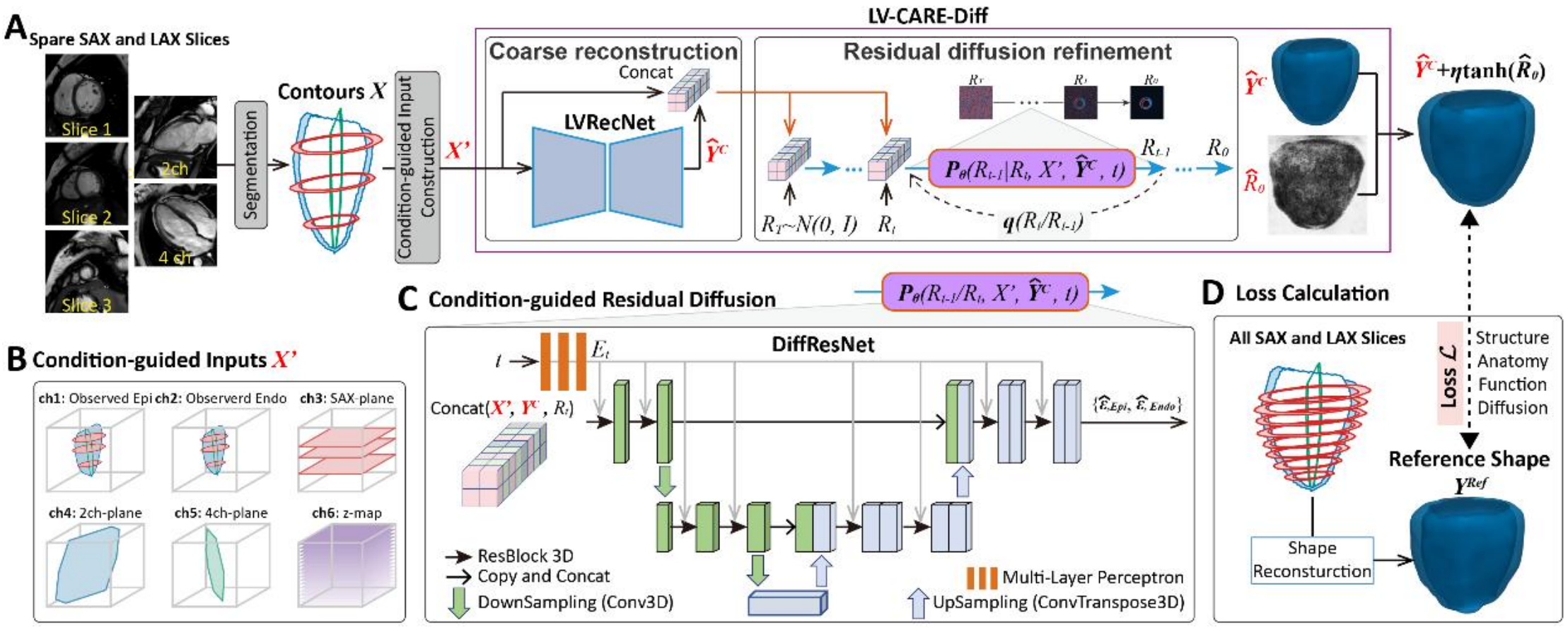

**Figure 1. Overview of LV-CARE-Diff. A:** Study aim and network architecture of LV-CARE-Diff. LV-CARE-Diff employs a coarse-to-fine, end-to-end framework to reconstruct the complete LV shape from ultra-sparse cine acquisitions (three SAX + two LAX slices), enabling acceleration of routine cine examinations that otherwise require a full SAX stack. LV-CARE-Diff comprises two substeps: a 3DUNet first predicts a coarse LV shape, which is subsequently refined by a condition-guided residual diffusion model. **B:** The contours and spatial positions of the three SAX and two LAX slices are encoded as a multi-channel condition tensor *X'*, which provides LV shape context to guide the full LV-CARE-Diff reconstruction. **C:** The time-conditioned 3D ResUNet employed for denoising within the diffusion model. **D:** Multiple loss functions constraining reconstruction fidelity across structural, anatomical, functional, and diffusion denoising objectives.

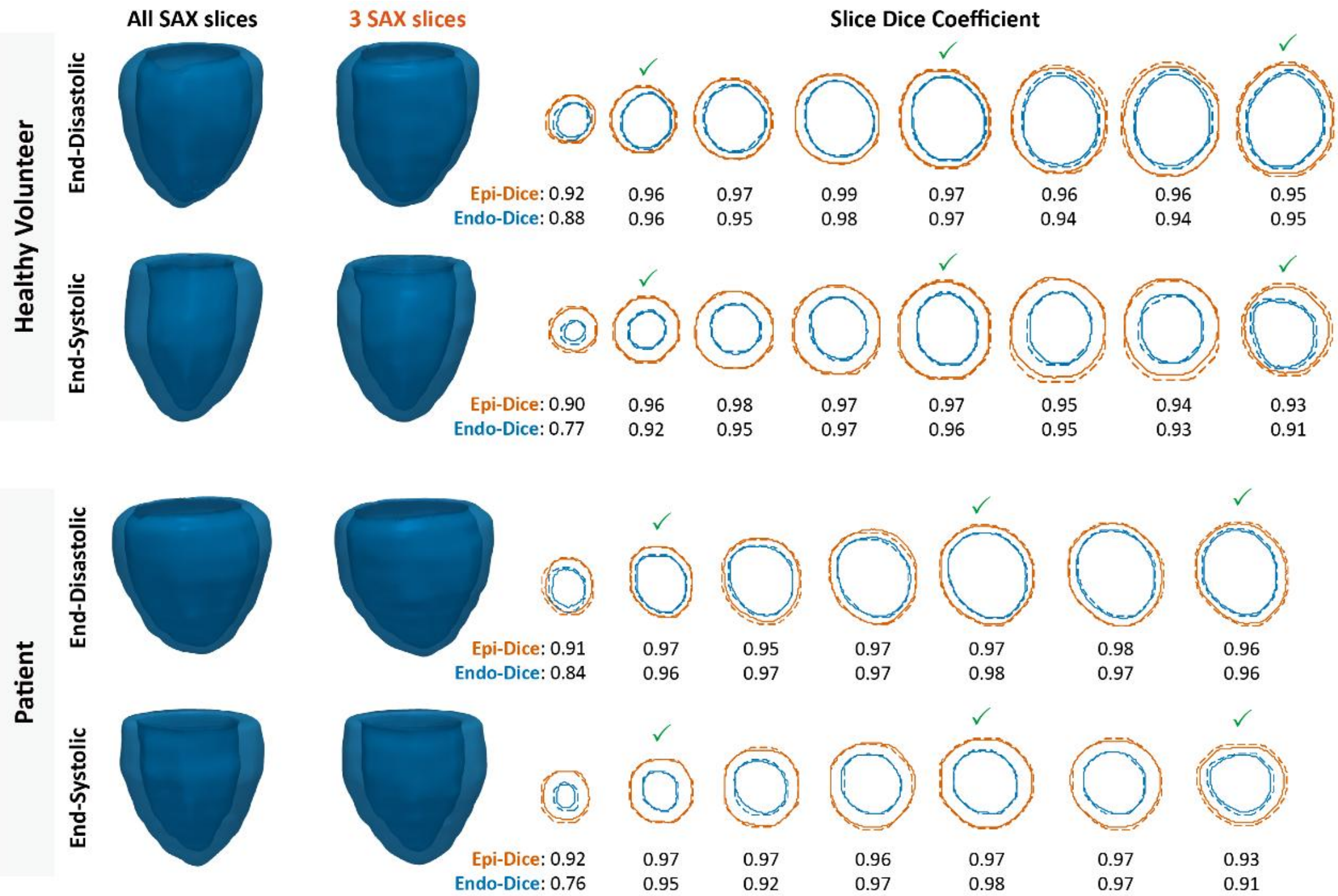

**Figure 2.** Comparison of left ventricular volumes reconstructed from three short-axis and two long-axis slices of a healthy volunteer and a patient volunteer using LV-CARE-Diff. The Dice coefficient of epicardial and endocardial contours was computed between cross-sections retrospectively derived from the LV-CARE-Diff-reconstructed volume and the originally acquired slices. Slices marked with a green check mark represent the input SAX slices.

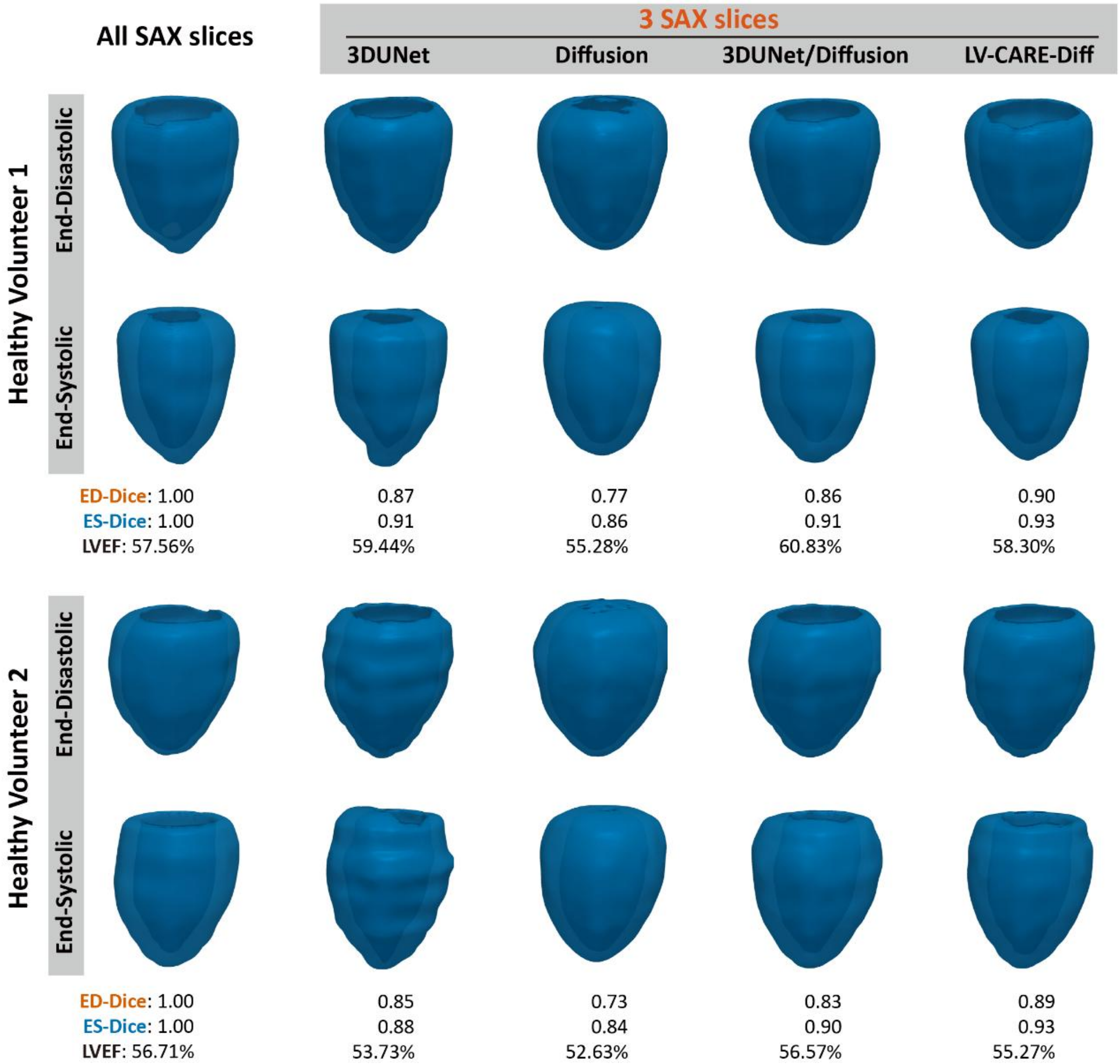

**Figure 3**. Visual comparison of left ventricular volumes reconstructed from three short-axis and two long-axis slices of two healthy volunteers using different methods. The Dice coefficient for the entire LV volume and the LV functional indices are reported below.

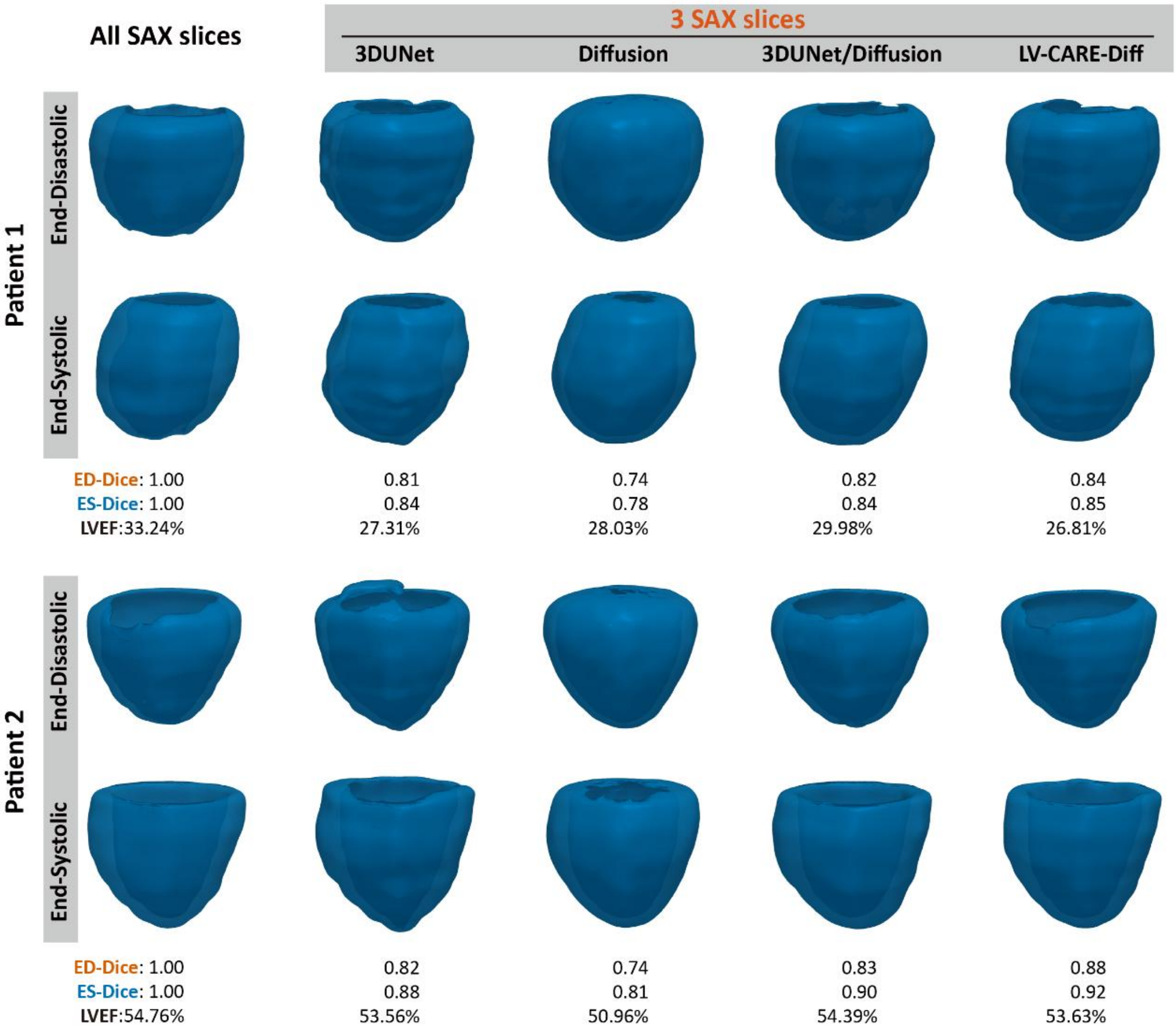

**Figure 4.** Visual comparison of left ventricular volumes reconstructed from three short-axis and two long-axis slices for two patients using different methods. The Dice coefficient for the entire LV volume and the LV functional indices are reported below.

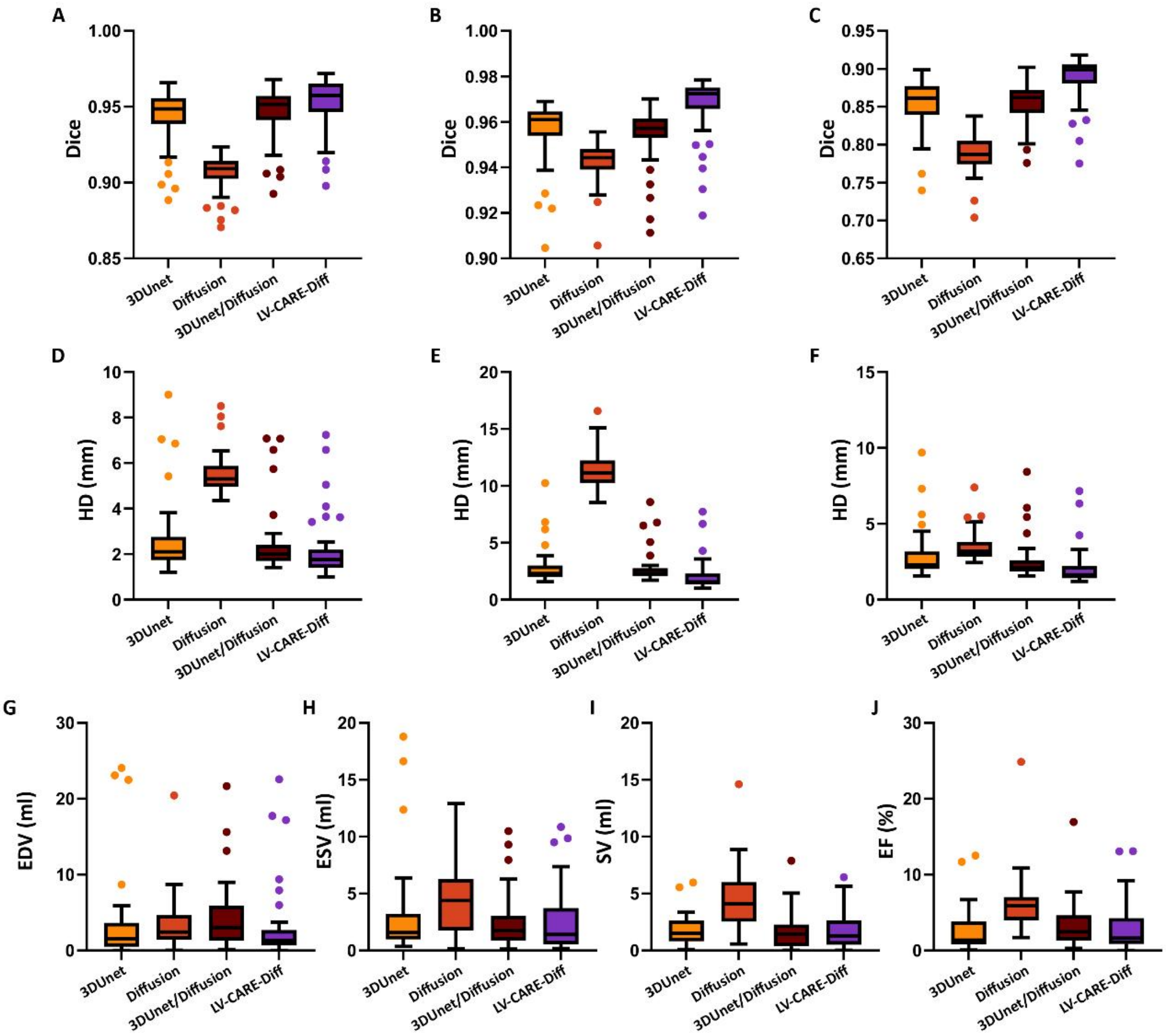

**Figure 5.** Comparison of Dice coefficients and 95% Hausdorff distance (HD) for reconstructed LV shape across four deep learning models, and the corresponding MAE for LV functional index quantification.

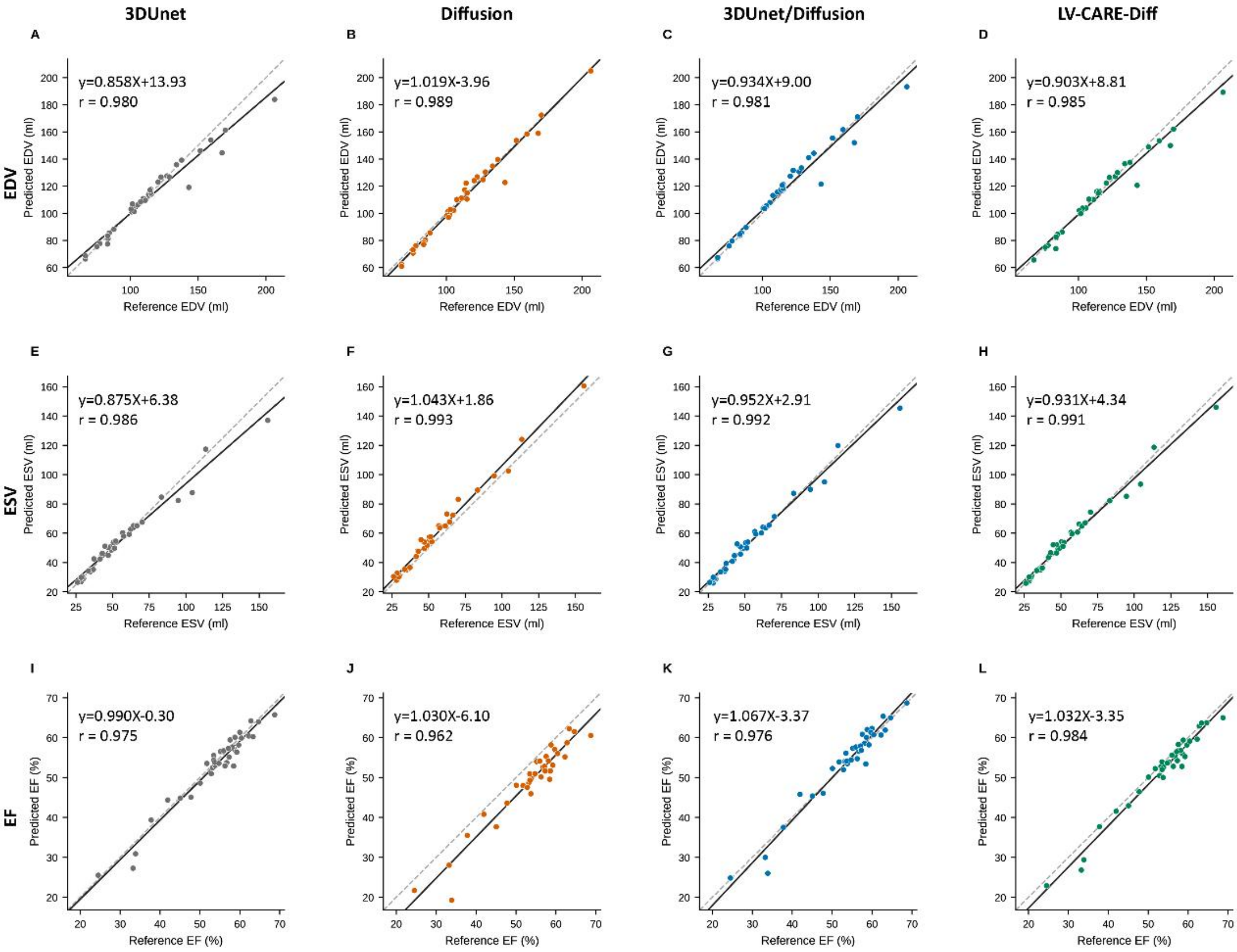

**Figure 6.** Correlation between LV functional indices estimated from all slices and ultra-sparse slices using deep-learning reconstruction.

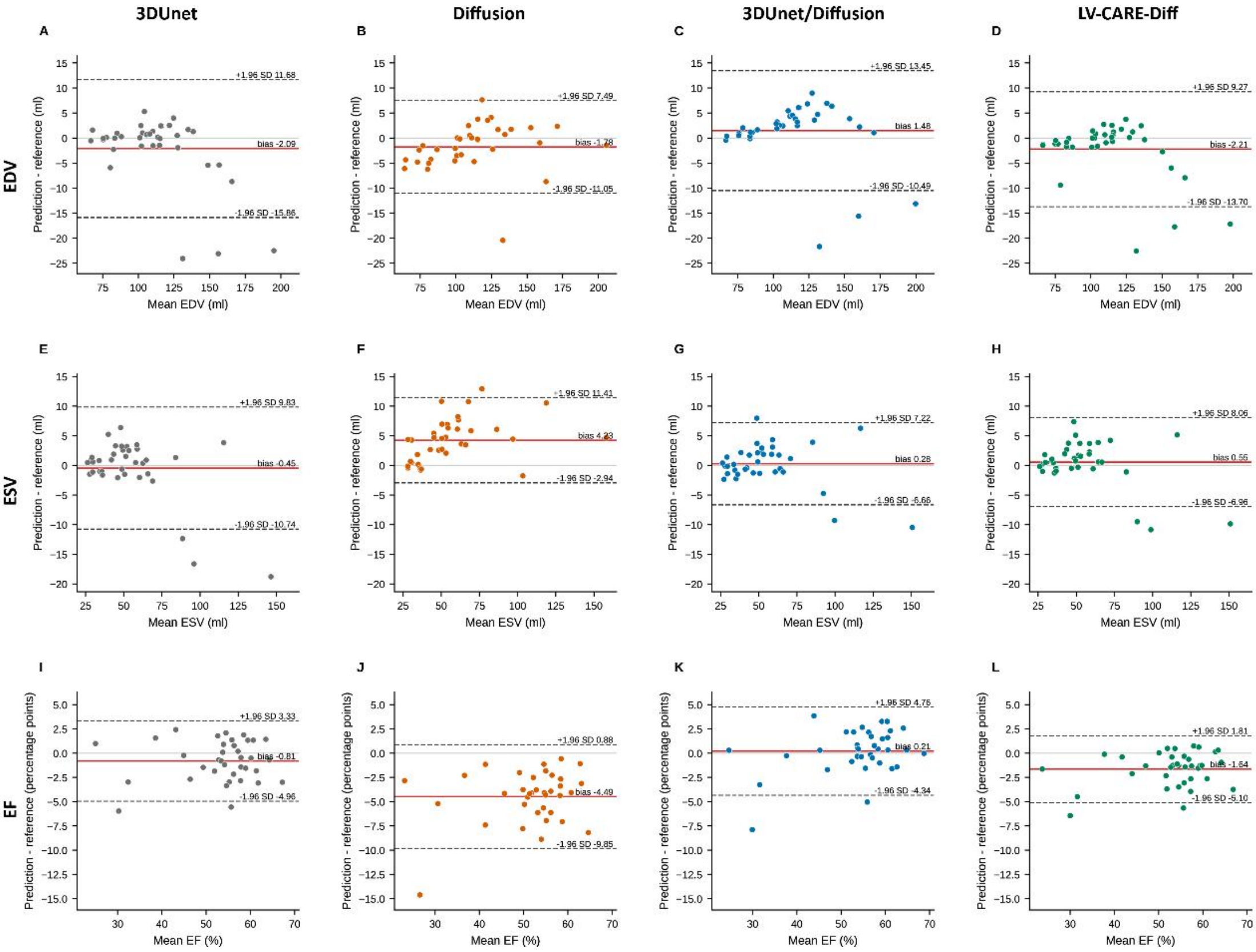

**Figure 7.** Bland-Altman plots for examining the agreement in LV functional index quantification between the reference standard derived from all SAX slices and the estimates obtained from ultra-sparse SAX slices.

**Table 1.** Quantitative performance comparison of different deep-learning models (N=36)

| | **LV shape reconstruction** | | | | | |
|---|---|---|---|---|---|---|
| | **Dice** | | | **HD (mm)** | | |
| | **Endocardium** | **Epicardium** | **Myocardium** | **Endocardium** | **Epicardium** | **Myocardium** |
| 3DUNet | 0.943 ± 0.020 | 0.956 ± 0.015 | 0.854 ± 0.035 | 2.708 ± 1.734 | 2.843 ± 1.722 | 2.956 ± 1.659 |
| Diffusion | 0.906 ± 0.013 | 0.942 ± 0.009 | 0.787 ± 0.026 | 5.572 ± 0.946 | 11.399 ± 1.776 | 3.467 ± 1.007 |
| 3DUnet/Diffusion | 0.946 ± 0.019 | 0.953 ± 0.013 | 0.854 ± 0.028 | 2.514 ± 1.558 | 2.770 ± 1.534 | 2.613 ± 1.393 |
| LV-CARE-Diff | **0.952 ± 0.019** | **0.967 ± 0.014** | **0.886 ± 0.033** | **2.257 ± 1.456** | **2.101 ± 1.476** | **2.155 ± 1.327** |

| | **LV function quantification MAE** | | | |
|---|---|---|---|---|
| | **EDV (ml)** | **ESV (ml)** | **SV (ml)** | **EF (%)** |
| 3DUNet | 3.74±6.28 | 3.13±4.21 | **2.76±2.99** | 1.82±1.32 |
| Diffusion | 3.48±3.63 | 4.47±3.36 | 6.11±3.95 | 4.49±2.74 |
| 3DUnet/Diffusion | 4.31±4.52 | **2.48±2.51** | 3.25±3.06 | **1.65±1.62** |
| LV-CARE-Diff | **3.29±5.31** | 2.59±2.85 | 2.97±3.35 | 1.80±1.60 |

**Table 2.** Impact of cardiac phase (N=36)

| | Dice | | | HD (mm) | | |
|---|---|---|---|---|---|---|
| | **Endocardium** | **Epicardium** | **Myocardium** | **Endocardium** | **Epicardium** | **Myocardium** |
| **End-Diastolic phase** | **0.960±0.023** | **0.969±0.017** | 0.865±0.040 | **2.255 ± 1.731** | 2.164 ± 1.801 | **2.135 ± 1.585** |
| **End-Systolic phase** | 0.944±0.019 | 0.965±0.012 | **0.906±0.029** | 2.258 ± 1.314 | **2.038 ± 1.213** | 2.174 ± 1.119 |

**Table 3.** Performance Comparison Between Healthy Volunteers and Patients

| | LV shape reconstruction | | | | | |
|---|---|---|---|---|---|---|
| | Dice | | | HD (mm) | | |
| | Endocardium | Epicardium | Myocardium | Endocardium | Epicardium | Myocardium |
| Healthy volunteers (N=25) | **0.960 ± 0.008** | **0.973 ± 0.005** | **0.899 ± 0.015** | **1.638 ± 0.360** | **1.502 ± 0.379** | **1.613 ± 0.327** |
| Patients (N=11) | 0.934 ± 0.025 | 0.954 ± 0.019 | 0.854 ± 0.042 | 3.663 ± 1.995 | 3.461 ± 2.086 | 3.385 ± 1.872 |

| | LV function quantification MAE | | | |
|---|---|---|---|---|
| | EDV (ml) | ESV (ml) | SV (ml) | EF (%) |
| Healthy volunteers (N=25) | **1.47±1.32** | **1.94±1.79** | **2.10±1.65** | **1.71±1.34** |
| Patients (N=11) | 7.41±8.23 | 4.05±4.16 | 4.97±5.13 | 2.01±2.13 |

**Table 4.** The impact of condition-guided input with/without LAX slice (n=36)

| | LV shape reconstruction | | | | | |
|---|---|---|---|---|---|---|
| | Dice | | | HD (mm) | | |
| | Endocardium | Epicardium | Myocardium | Endocardium | Epicardium | Myocardium |
| 2ch+4ch | **0.952 ± 0.019** | **0.967 ± 0.014** | **0.886 ± 0.033** | **2.257 ± 1.456** | **2.101 ± 1.476** | **2.155 ± 1.327** |
| 4ch | 0.950 ± 0.019 | 0.963 ± 0.015 | 0.876 ± 0.038 | 2.380 ± 1.366 | 2.450 ± 1.523 | 2.376 ± 1.271 |
| 2ch | 0.950 ± 0.019 | 0.963 ± 0.015 | 0.879 ± 0.035 | 2.589 ± 1.827 | 2.729 ± 2.315 | 2.375 ± 1.442 |

| | LV function quantification | | | |
|---|---|---|---|---|
| | EDV (ml) | ESV (ml) | SV (ml) | EF (%) |
| 2ch+4ch | 3.29±5.31 | 2.59±2.85 | **2.97±3.35** | **1.80±1.60** |
| 4ch | **3.26 ± 4.77** | **2.48 ± 2.88** | 3.21 ± 3.04 | 1.99 ± 1.59 |
| 2ch | 3.77 ± 6.06 | 2.56 ± 2.98 | 3.19 ± 3.79 | 1.81 ± 1.65 |

**Table 5.** Impact of the number of SAX slices (n=36)

| | **LV shape reconstruction** | | | | | |
|---|---|---|---|---|---|---|
| | **Dice** | | | **HD (mm)** | | |
| | **Endocardium** | **Epicardium** | **Myocardium** | **Endocardium** | **Epicardium** | **Myocardium** |
| 3 | 0.952 ± 0.019 | 0.967 ± 0.014 | 0.886 ± 0.033 | 2.257 ± 1.456 | 2.101 ± 1.476 | 2.155 ± 1.327 |
| 4 | 0.966 ± 0.014 | 0.976 ± 0.009 | 0.914 ± 0.029 | 1.613 ± 0.775 | 1.567 ± 0.913 | 1.615 ± 0.881 |
| 5 | **0.971 ± 0.010** | **0.980 ± 0.008** | **0.928 ± 0.025** | **1.502 ± 0.704** | **1.366 ± 0.715** | **1.460 ± 0.704** |

| | **LV function quantification** | | | |
|---|---|---|---|---|
| | **EDV (ml)** | **ESV (ml)** | **SV (ml)** | **EF (%)** |
| 3 | 3.29±5.31 | 2.59±2.85 | 2.97±3.35 | 1.80±1.60 |
| 4 | 2.01±3.35 | 1.36±2.07 | 1.68±1.59 | 0.91±0.64 |
| 5 | **1.76±2.52** | **1.13±1.50** | **1.65±1.56** | **0.86±0.66** |

**Table 6**. Impact of Slice Location Shift (n=36)

| | LV shape reconstruction | | | | | |
|---|---|---|---|---|---|---|
| | Dice | | | HD (mm) | | |
| | Endocardium | Epicardium | Myocardium | Endocardium | Epicardium | Myocardium |
| **Precalculated** | 0.952 ± 0.019 | **0.967 ± 0.014** | **0.886 ± 0.033** | **2.257 ± 1.456** | **2.101 ± 1.476** | **2.155 ± 1.327** |
| **Shifted** | 0.939 ± 0.024 | 0.953 ± 0.019 | 0.856 ± 0.043 | 3.320 ± 1.887 | 3.397 ± 1.902 | 3.082 ± 1.587 |

| | LV function quantification | | | |
|---|---|---|---|---|
| | EDV (ml) | ESV (ml) | SV (ml) | EF (%) |
| **Precalculated** | **3.29±5.31** | **2.59±2.85** | **2.97±3.35** | **1.80±1.60** |
| **Shifted** | 8.35±9.25 | 3.73±4.57 | 7.42±8.51 | 3.60±4.06 |

# Supplementary information

**Table S1.** Left ventricular functional quantitative results, relative errors, and paired t-test p-values for comparison across deep learning models.

| | Metric | All SAX slices (Reference) | 3 SAX slices (Prediction) | Difference | Relative error (%) | P |
|---|---|---|---|---|---|---|
| **3DUNet** | **EDV (mL)** | 113.10 ± 31.30 | 111.01 ± 27.41 | -2.09 ± 7.03 | **-1.19 ± 4.56** | 0.083 |
| | **ESV (mL)** | 54.58 ± 27.21 | 54.13 ± 24.13 | -0.45 ± 5.25 | **0.67 ± 6.43** | 0.606 |
| | **SV (mL)** | 58.52 ± 11.12 | 56.88 ± 11.40 | -1.64 ± 3.74 | -2.71 ± 6.71 | 0.013 |
| | **EF (%)** | 53.67 ± 9.39 | 52.86 ± 9.53 | -0.81 ± 2.12 | -1.56 ± 4.67 | 0.027 |
| **Diffusion** | **EDV (mL)** | 113.10 ± 31.30 | 111.32 ± 32.25 | -1.78 ± 4.73 | -1.94 ± 4.11 | 0.030 |
| | **ESV (mL)** | 54.58 ± 27.21 | 58.81 ± 28.60 | 4.23 ± 3.66 | 7.78 ± 6.64 | <0.001 |
| | **SV (mL)** | 58.52 ± 11.12 | 52.50 ± 11.63 | -6.01 ± 4.11 | -10.55 ± 8.24 | <0.001 |
| | **EF (%)** | 53.67 ± 9.39 | 49.19 ± 10.04 | -4.49 ± 2.74 | -8.87 ± 7.10 | <0.001 |
| **3DUNet/Diffusion** | **EDV (mL)** | 113.10 ± 31.30 | 114.58 ± 29.78 | 1.48 ± 6.11 | 1.66 ± 4.20 | 0.154 |
| | **ESV (mL)** | 54.58 ± 27.21 | 54.86 ± 26.11 | 0.28 ± 3.54 | 0.93 ± 5.34 | 0.635 |
| | **SV (mL)** | 58.52 ± 11.12 | 59.72 ± 12.83 | 1.20 ± 4.33 | **1.86 ± 8.05** | 0.105 |
| | **EF (%)** | 53.67 ± 9.39 | 53.88 ± 10.26 | 0.21 ± 2.32 | **0.09 ± 5.45** | 0.593 |
| **LV-CARE-Diff** | **EDV (mL)** | 113.10 ± 31.30 | 110.89 ± 28.69 | -2.21 ± 5.86 | -1.60 ± 4.02 | 0.030 |
| | **ESV (mL)** | 54.58 ± 27.21 | 55.13 ± 25.54 | 0.55 ± 3.83 | 1.93 ± 5.31 | 0.395 |
| | **SV (mL)** | 58.52 ± 11.12 | 55.76 ± 11.70 | -2.76 ± 3.53 | -4.86 ± 6.40 | <0.001 |
| | **EF (%)** | 53.67 ± 9.39 | 52.03 ± 9.84 | -1.64 ± 1.76 | -3.37 ± 4.19 | <0.001 |

**Table S2.** Left ventricular functional quantitative results, relative errors, and paired t-test p-values for statistical comparison between end-diastolic and end-systolic cardiac phases.

| | Metric | All SAX slices (Reference) | 3 SAX slices (Prediction) | Difference | Relative error (%) | P |
|---|---|---|---|---|---|---|
| **End-Diastolic phase** | **EDV (mL)** | 113.10 ± 31.30 | 110.89 ± 28.69 | -2.21 ± 5.86 | -1.60 ± 4.02 | 0.030 |
| **End-Systolic phase** | **ESV (mL)** | 54.58 ± 27.21 | 55.13 ± 25.54 | 0.55 ± 3.83 | 1.93 ± 5.31 | 0.395 |

**Table S3.** Left ventricular functional quantitative results, relative errors, and paired t-test p-values for statistical comparison between healthy volunteers and patients.

| | Metric | All SAX slices (Reference) | 3 SAX slices (Prediction) | Difference | Relative error (%) | P |
|---|---|---|---|---|---|---|
| **Healthy volunteers (N=25)** | **EDV (mL)** | 103.88 ± 24.23 | 103.59 ± 24.25 | -0.29 ± 1.98 | -0.35 ± 1.66 | 0.472 |
| | **ESV (mL)** | 46.51 ± 14.53 | 48.06 ± 15.09 | 1.54 ± 2.16 | 3.19 ± 4.53 | 0.002 |
| | **SV (mL)** | 57.36 ± 10.85 | 55.53 ± 10.09 | -1.83 ± 1.95 | -3.06 ± 2.76 | <0.001 |
| | **EF (%)** | 55.88 ± 4.49 | 54.36 ± 4.52 | -1.52 ± 1.56 | -2.71 ± 2.75 | <0.001 |
| **Patients (N=11)** | **EDV (mL)** | 134.05 ± 36.39 | 127.47 ± 32.14 | -6.58 ± 8.97 | -4.44 ± 6.09 | 0.035 |
| | **ESV (mL)** | 72.92 ± 39.40 | 71.21 ± 36.43 | -1.71 ± 5.67 | -0.94 ± 6.03 | 0.341 |
| | **SV (mL)** | 61.14 ± 11.81 | 56.26 ± 15.32 | -4.87 ± 5.23 | -8.96 ± 9.92 | 0.011 |
| | **EF (%)** | 48.67 ± 14.84 | 46.74 ± 15.67 | -1.93 ± 2.21 | -4.89 ± 6.31 | 0.016 |

**Table S4.** Left ventricular functional quantitative results, relative errors, and paired t-test p-values for investigating the impact of LAX slice input configuration.

| | Metric | All SAX slices (Reference) | 3 SAX slices (Prediction) | Difference | Relative error (%) | P |
|---|---|---|---|---|---|---|
| 2ch+4ch | **EDV (mL)** | 113.10 ± 31.30 | 110.89 ± 28.69 | -2.21 ± 5.86 | -1.60 ± 4.02 | 0.030 |
| | **ESV (mL)** | 54.58 ± 27.21 | 55.13 ± 25.54 | 0.55 ± 3.83 | 1.93 ± 5.31 | 0.395 |
| | **SV (mL)** | 58.52 ± 11.12 | 55.76 ± 11.70 | -2.76 ± 3.53 | -4.86 ± 6.40 | <0.001 |
| | **EF (%)** | 53.67 ± 9.39 | 52.03 ± 9.84 | -1.64 ± 1.76 | -3.37 ± 4.19 | <0.001 |
| 4ch | **EDV (mL)** | 113.10 ± 31.30 | 110.52 ± 29.04 | -2.57 ± 5.18 | -2.00 ± 3.63 | 0.005 |
| | **ESV (mL)** | 54.58 ± 27.21 | 54.96 ± 25.24 | 0.38 ± 3.81 | 1.69 ± 5.49 | 0.558 |
| | **SV (mL)** | 58.52 ± 11.12 | 55.57 ± 11.32 | -2.95 ± 3.30 | -5.09 ± 5.69 | <0.001 |
| | **EF (%)** | 53.67 ± 9.39 | 52.01 ± 9.48 | -1.66 ± 1.94 | -3.19 ± 3.97 | <0.001 |
| 2ch | **EDV (mL)** | 113.10 ± 31.30 | 110.41 ± 27.96 | -2.69 ± 6.62 | -1.92 ± 4.42 | 0.020 |
| | **ESV (mL)** | 54.58 ± 27.21 | 54.40 ± 25.37 | -0.18 ± 3.95 | 0.52 ± 5.72 | 0.788 |
| | **SV (mL)** | 58.52 ± 11.12 | 56.00 ± 11.98 | -2.51 ± 4.28 | -4.42 ± 7.65 | 0.001 |
| | **EF (%)** | 53.67 ± 9.39 | 52.48 ± 10.12 | -1.19 ± 2.15 | -2.65 ± 5.00 | 0.002 |

**Table S5.** Left ventricular functional quantitative results, relative errors, and paired t-test p-values for investigating the impact of SAX slices.

| | Metric | All SAX slices (Reference) | Prediction | Difference | Relative error (%) | P |
|---|---|---|---|---|---|---|
| **3 SAX Slices** | **EDV (mL)** | 113.10 ± 31.30 | 110.89 ± 28.69 | -2.21 ± 5.86 | -1.60 ± 4.02 | 0.030 |
| | **ESV (mL)** | 54.58 ± 27.21 | 55.13 ± 25.54 | 0.55 ± 3.83 | 1.93 ± 5.31 | 0.395 |
| | **SV (mL)** | 58.52 ± 11.12 | 55.76 ± 11.70 | -2.76 ± 3.53 | -4.86 ± 6.40 | <0.001 |
| | **EF (%)** | 53.67 ± 9.39 | 52.03 ± 9.84 | -1.64 ± 1.76 | -3.37 ± 4.19 | <0.001 |
| **4 SAX Slices** | **EDV (mL)** | 113.10 ± 31.30 | 111.81 ± 31.30 | -1.29 ± 3.70 | -1.17 ± 3.10 | 0.044 |
| | **ESV (mL)** | 54.58 ± 27.21 | 54.49 ± 26.87 | -0.09 ± 2.48 | 0.14 ± 4.02 | 0.822 |
| | **SV (mL)** | 58.52 ± 11.12 | 57.32 ± 11.33 | -1.19 ± 1.99 | -2.15 ± 3.47 | <0.001 |
| | **EF (%)** | 53.67 ± 9.39 | 53.13 ± 9.20 | -0.55 ± 0.97 | -0.98 ± 1.93 | 0.002 |
| **5 SAX Slices** | **EDV (mL)** | 113.10 ± 31.30 | 112.13 ± 31.52 | -0.96 ± 2.93 | -0.94 ± 2.33 | 0.056 |
| | **ESV (mL)** | 54.58 ± 27.21 | 54.67 ± 27.11 | 0.09 ± 1.88 | 0.31 ± 3.00 | 0.775 |
| | **SV (mL)** | 58.52 ± 11.12 | 57.46 ± 11.28 | -1.05 ± 2.02 | -1.88 ± 3.33 | 0.004 |
| | **EF (%)** | 53.67 ± 9.39 | 53.15 ± 9.22 | -0.52 ± 0.96 | -0.95 ± 2.05 | 0.002 |

**Table S6.** Left ventricular functional quantitative results, relative errors, and paired t-test p-values for investigating the impact of SAX slice location shift.

| | Metric | All SAX slices (Reference) | 3 SAX slices (Prediction) | Difference | Relative error (%) | P |
|---|---|---|---|---|---|---|
| Precalculated | EDV (mL) | 113.10 ± 31.30 | 110.89 ± 28.69 | -2.21 ± 5.86 | -1.60 ± 4.02 | **0.030** |
| | ESV (mL) | 54.58 ± 27.21 | 55.13 ± 25.54 | 0.55 ± 3.83 | 1.93 ± 5.31 | 0.395 |
| | SV (mL) | 58.52 ± 11.12 | 55.76 ± 11.70 | -2.76 ± 3.53 | -4.86 ± 6.40 | **<0.001** |
| | EF (%) | 53.67 ± 9.39 | 52.03 ± 9.84 | -1.64 ± 1.76 | -3.37 ± 4.19 | **<0.001** |
| Shifted | EDV (mL) | 113.10 ± 31.30 | 107.69 ± 26.10 | -5.41 ± 11.28 | -3.67 ± 8.53 | **0.007** |
| | ESV (mL) | 54.58 ± 27.21 | 54.62 ± 24.77 | 0.04 ± 5.93 | 1.41 ± 8.06 | 0.971 |
| | SV (mL) | 58.52 ± 11.12 | 53.07 ± 9.83 | -5.44 ± 9.93 | -7.89 ± 15.81 | **0.002** |
| | EF (%) | 53.67 ± 9.39 | 51.11 ± 10.38 | -2.56 ± 4.80 | -4.91 ± 10.97 | **0.003** |

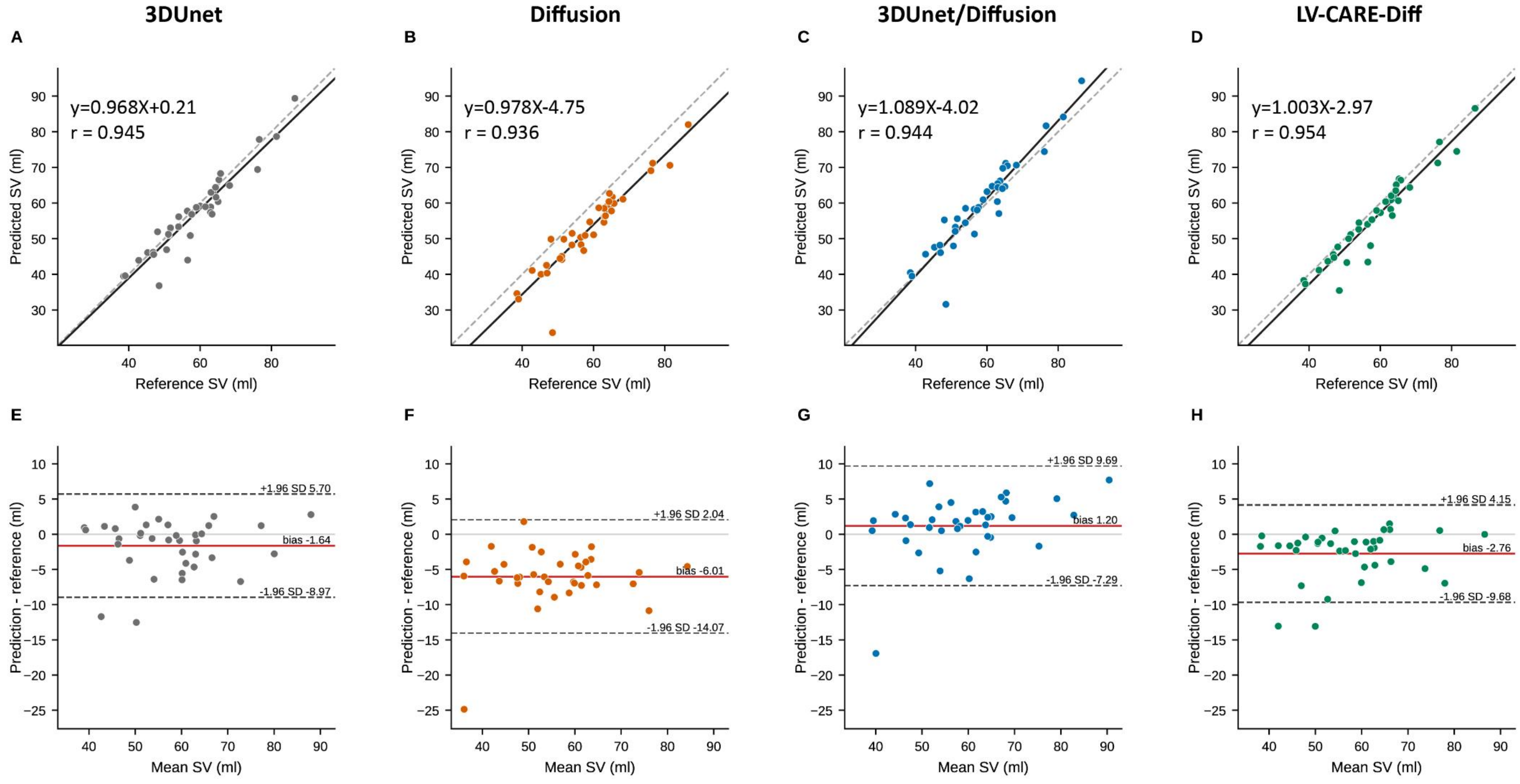

**Figure S1.** Correlation and agreement in SV between the reference standard derived from all SAX slices and estimates obtained from retrospectively under-sampled ultra-sparse SAX slices using LV-CARE-Diff.